\documentstyle[12pt]{article}
\def\be{\begin{equation}}
\def\ee{\end{equation}}
\def\bea{\begin{eqnarray}}
\def\eea{\end{eqnarray}}

\def\l{\lambda}
\def\t{\tau}

\def\pa{\partial}
\def\e{\epsilon}
\def\d{\delta}

\author{Ines Quandt and Hans-J\"urgen Schmidt}

\title{The Newtonian limit of fourth and higher order 
gravity\footnote{Lecture read at the seminar on 
Physical Implications of Gravitational Field Equations of 
 Higher Order, Potsdam, 3 - 5 October 1990}}

\date{}
\begin{document}
\maketitle

\centerline{Universit\"at Potsdam, Institut f\"ur Mathematik, Am
Neuen Palais 10} 
 \centerline{D-14469~Potsdam, Germany,  E-mail:
 hjschmi@rz.uni-potsdam.de}

\begin{abstract}
\noindent
We consider the Newtonian limit of the theory based on the Lagrangian 
$$
{\cal L} = \left( R + \sum_{k=0}^p \, a_k \, R \Box^k R 
 \right) \sqrt{-g} \, .
$$
The gravitational potential of
a point mass turns out to be a 
combination of Newtonian and Yukawa terms. 
For sixth-order gravity ($p = 1$) the coefficients are calculated
explicitly. For general $p$ one gets 
$$
\Phi  = m/r \, \left( 1 +  \sum_{i=0}^p \,   c_i  \exp (-r/l_i) \right)
$$
with $  \sum_{i=0}^p \,   c_i = 1/3$.  
Therefore, the potential is always unbounded near
the origin.    

\medskip

Wir betrachten den Newtonschen Grenzwert 
der durch den Lagrangian
$$
{\cal L} = \left( R + \sum_{k=0}^p \, a_k \, R \Box^k R 
 \right) \sqrt{-g} 
$$
beschriebenen Theorie. 
Das Gravitationspotential einer Punktmasse hat die Gestalt von 
Kombinationen von Newton- und Yukawa-Termen. F\"ur 
Gravitationsgleichungen sechster Ordnung ($p = 1$) wird das 
Fern\-feld von 
Punktmassen explizit angegeben. 
F\"ur allgemeine $p$ wird die Summe der
Koeffizienten vor den Yukawa-Termen durch 
 $  \sum_{i=0}^p \,   c_i = 1/3$
beschrieben.  Also ist das Potential stets unbeschr\"ankt.  


\noindent 
AAA subject classification: 066
\end{abstract}

\section{Introduction}

One-loop quantum corrections
 to the Einstein equation can be described by curvature-squared terms and 
lead to fourth-order
gravitational field equations; their Newtonian limit is described by a 
potential ``Newton + one Yukawa term", cf. e.g.
STELLE (1978) and TEYSSANDIER (1989).
 A Yukawa potential has the form $\exp (-r/l)/r$  and was originally 
used by YUKAWA
(1935) to describe the meson field.

\bigskip

Higher-loop quantum corrections to the Einstein equation are expected to 
contain terms of the type $
R \Box^k R$  in the Lagrangian, which leads to a 
gravitational field equation of order $2k + 4$, cf.
 GOTTL\"OBER et al. (1990). 
Some preliminary results to this type of equations are already due 
to BUCHDAHL (1951).

\bigskip

In the present paper we deduce the 
Newtonian limit following from this higher order field equation. The 
Newtonian 
limit of General Relativity Theory is the usual Newtonian theory, cf.  
e.g. DAUTCOURT (1964) or STEPHANI (1977). From the general 
structure of the linearized higher-order field equation, 
cf.  SCHMIDT (1990), one can expect that for this higher-order 
gravity the far field of the point mass in the Newtonian limit 
is the Newtonian  potential plus a sum of different Yukawa terms. 
And just this form is that one discussed 
in connection with the fifth force, 
cf. GERBAL and SIROUSSE-ZIA (1989),  
STACEY et al. (1981) and SANDERS (1984, 1986).

\bigskip

Here we are interested in the details of this connection 
between higher-order gravity and the lengths and coefficients in 
the corresponding Yukawa terms.

\section{Lagrangian and field equation}

Let us start with the Lagrangian
\be
{\cal L} = \left( R + \sum_{k=0}^p \, a_k \, R \Box^k R 
 \right) \cdot \sqrt{-g} 
\, ,  \qquad a_p \ne 0 \, .
\ee
In our considerations we will assume 
that for the gravitational  constant $G$
 and for the speed of  light $c$ 
it holds $G = c = 1$. This only means a special 
choice of units. In eq. (1), $R$ denotes the curvature scalar, 
$\Box$ the D'Alembertian, and $g$ the determinant of the metric. 
Consequently, the coefficient $a_k$ 
 has the dimension ``length to the power $2k + 2$".

\bigskip

The starting point for the deduction of the field 
equation is the principle of minimal 
action. A necessary condition for it is the 
stationarity of the action: 
$$
- \frac{\d {\cal L}}{\d g_{ij}} = 
8 \pi T^{ij} \sqrt{-g} \, ,
$$
 where $T^{ij}$ denotes 
the energy-momentum tensor. The explicit
equations for 
$$
P^{ij} \sqrt{-g}= - \frac{\d {\cal L}}{\d g_{ij}}
$$
    are given by SCHMIDT (1990). Here we only need the 
linearized field equation. It reads, cf. GOTTL\"OBER
 et al. (1990)   
\be
P^{ij}
\equiv R^{ij} - \frac{R}{2} \, 
 g^{ij} + 2 \sum_{k=0}^p \, a_k  [
 g^{ij} \Box^{k+1} \, R - \Box^k
 R^{\, ; \, ij} ] = 8 \pi T^{ij}\, , 
\ee
and for the trace it holds:
\be
g_{ij} \cdot P^{ij} =  -  \frac{n-1}{2}
 R + 2n   
\sum_{k=0}^p \, a_k [
 g^{ij} \Box^{k+1} \, R ]
 = 8 \pi T \, .
\ee
$n$ is the number of spatial dimensions; 
the most important application is of cause $n = 3$.
 From now on we put $n = 3$.

\section{The Newtonian limit}

The Newtonian limit is the weak-field static  
approximation. So we use the linearized 
field equation and insert a static metric and an energy-momentum tensor
\be
T_{ij} = \d^0_i \ \, \d^0_j \, \rho \, , \qquad \rho  \ge  0
\ee
into eq. (2).

\bigskip

Without proof we mention that the metric 
can be brought into spatially 
conformally flat form and so we use
\bea 
g_{ij} = \eta_{ij} + f_{ij } \, ,
  \nonumber \\
\eta_{ij} = {\rm  diag} (1, \,  -1, \,  -1, \,  -1)
 \qquad {\rm and}
 \nonumber \\
f_{ij} = {\rm  diag} (-2\Phi, \, -2\Psi , \,  -2\Psi , \,  -2\Psi ) \, .
 \nonumber 
\eea
So the metric equals
\be
ds^2 = (1 - 2\Phi) dt^2 - (1 + 2\Psi ) (dx^2 + dy^2 + dz^2) \, ,  
\ee
where $\Phi$ and $\Psi$  depend on $x$, $y$ and $z$.

\bigskip

Linearization means that the metric $g_{ij}$
 has only a small difference to  $\eta_{ij}$; 
quadratic expressions in  $f_{ij}$ and its 
partial derivatives are neglected.

\bigskip

We especially consider the case of a point mass. In this case it holds: 
$\Phi  = \Phi(r)$, $ \Psi  = \Psi (r)$,
 with 
$$
r = \sqrt{x^2+y^2 +z^2} \, , 
$$
 because of spherical  symmetry and $\rho = m \d$.  
Using these properties, we deduce the field equation
 and discuss the existence 
of solutions of the above mentioned type.

\bigskip

At first we make some helpful general considerations: The functions 
$\Phi$ and $\Psi $ 
are determined
 by eq. (2) for $i = j = 0$ and the trace of eq. (2). If these 
two equations hold, then
 all other component-equations are automatically 
satisfied. For the 00-equation we need $R_{00}$:
\be
R_{00} = - \Delta \Phi \, .
\ee
Here is as usual
$$
\Delta  =
 \frac{\pa^2}{\pa x^2} +  \frac{\pa^2}{\pa y^2}
 +  \frac{\pa^2}{\pa z^2} \, . 
$$
For the inverse metric we get
$$
g^{ij} = {\rm diag} \left( 1/
 (1 - 2\Phi) , \,   - 1/ (1 + 2\Psi ) , \,   - 
1/ (1 + 2\Psi ) , \,   - 1/ (1 + 2\Psi )
 \right) 
$$
and $ 1/
 (1 - 2\Phi)
 = 1 + 2\Phi  + h(\Phi )$, where $h(\Phi )$ is 
quadratic in $\Phi$  and vanishes 
after linearization. So 
we get 
$$
g^{ij} = \eta^{ij} - f^{ij} \, .
$$
 In our coordinate system, $f^{ij}$
 equals $f_{ij}$  for all $i, j$. 
For the curvature scalar we get
\be
     R = 2(2 \Delta \Psi - \Delta \Phi )     \, .
\ee
Moreover, we need expressions of the type
 $\Box^k \, R$. $\Box R$ is 
defined by $ \Box R  = R_{\, ; \, ij} \, g^{ij}$,   
where ``$;$" 
denotes the covariant derivative. 
Remarks: Because of linearization we may replace
 the covariant derivative with 
the partial one. So we get
\be
 \Box^k \,  R = (~1)^k \, 2( 
- \Delta^{k+1}
\Phi  + 2 \Delta^{k+1} \Psi )    
\ee
and after some calculus
\be
          -8 \pi \rho  = \Delta  \Phi  + \Delta \Psi \,  .    
\ee
We use eq. (9) to eliminate $\Psi$ from the system. 
So we get
 by help
 of eq. (8) an
 equation relating $\Phi$
  and $\rho  = m \d$.
\be
     -4\pi \left( \rho + 8  \sum_{k=0}^p \, a_k
 (-1)^k  \Delta^{k+1} \,  \rho \right) 
 = \Delta 
\Phi  + 6  \sum_{k=0}^p \, a_k  (-1)^k
\Delta^{k+2} \Phi \, .
\ee
In spherical coordinates it holds
$$
\Delta
 \Phi = \frac{2}{r}
\Phi_{\, , \, r }  +\Phi_{\, , \, r r } \, ,  
$$
because $\Phi$  depends on the radial coordinate $r$ only.

\bigskip

We apply the following lemma: In the sense of distributions it holds
\be
\Delta \left(
\frac{1}{r} e^{-r/l} \right) 
= \frac{1}{rl^2}e^{-r/l} - 4 \pi \d \, . 
\ee
Now we are ready to solve the whole problem. We assume
$$
\Phi  = \frac{m}{r} \, \left( 1 +  \sum_{i=0}^q \,
   c_i  \exp (-r/l_i) \right) \, , \quad l_i > 0 \, .
$$
Without loss of generality we may assume $l_i \ne l_j$
 for $i \ne j$. Then eq. 
(10) together with the lemma eq. (11) yield
\bea
8\pi   \sum_{k=0}^p \, a_k
 (-1)^k  \Delta^{k+1} \,  \d =
  \sum_{i=0}^q \, 
\left(\frac{c_i}{t_i}
+ 6 \sum_{k=0}^p \, a_k
 (-1)^k  
\frac{c_i}{t_i^{k+2}}
\right)
\frac{1}{rl^2}e^{-r/l_i}
\nonumber \\
- 4 \pi   \sum_{i=0}^q 
\left(c_i
+ 6 \sum_{k=0}^p  a_k
 (-1)^k  
\frac{c_i}{t_i^{k+1}}
\right) \d 
 \nonumber \\
+ 24 \pi \sum_{k=0}^p
\sum_{j=k}^p \sum_{i=0}^p
 c_i a_j (-1)^{j+1} \frac{1}{t_i^{j-k}}
 \Delta^{k+1} \d 
\nonumber
\eea
where $t_i = l_i^2$; therefore also $t_i \ne t_j$ for $i\ne  j$.
This equation is equivalent to the system (12, 13, 14)
\bea
\sum_{i=0}^q c_i = 1/3
\\
\sum_{i=0}^q \frac{c_i}{t_i^s} = 0 \qquad s = 1, \dots p
\\
t_i^{p+1} + 6 \sum_{k=0}^p
a_k (-1)^k t_i^{p-k }
= 0 \qquad i=0, \dots q \, . 
\eea
From eq. (14) we see that the values 
$t_i$ represent $q + 1$ different solutions of one polynomial. This 
polynominal
 has the  degree $p + 1$. Therefore $q \le  p$.

\bigskip

Now we use eqs. (12) and (13). They can be written in matrix form as
$$
 \left( \begin{array}{c}
1 \dots  1\\ 
1/ t_0  \dots  1/t_q
 \\ \dots \\ 
1/ t_0^p  \dots  1/t_q^p
 \end{array} \right)  \cdot 
 \left( \begin{array}{c}
c_0 \\ c_1 \\ \dots \\ c_q \end{array} 
 \right) 
= \left( \begin{array}{c}
1/3 \\ 0 \\ \dots \\ 0 \end{array} 
\right)
$$
Here, the first $q + 1$ rows form a regular matrix (Vandermonde matrix). 
Therefore, we get
$$
1/t_i^j = 
\sum_{k=0}^q    \l_{jk} \, / \,  t_i^k
\qquad
 j = q + 1, \dots    p
$$
with certain coefficients $    \l_{jk}$
 i.e., the remaining 
rows depend on the first $q + 1$ ones. If $    \l_{j0} \ne 0$
 then the system has no
solution. So   $    \l_{j0}= 0$  for all $ q + 1 \le  j \le  p$. 
But for $q < p$ we would get 
$$
1/t_i^q = 
\sum_{k=1}^q    \l_{q+1\, k} \, / \,  t_i^{k-1} 
$$
and this is a contradiction
to the above stated regularity. 
Therefore $p$ equals $q$. The polynomial in (14) may be written as
$$
6 \cdot  \left( \begin{array}{c}
1 \  1/t_0 \dots  1/t_0^p\\ 
 \dots \\ 
1 \  1/ t_p  \dots  1/t_p^p
 \end{array} \right)  \cdot 
 \left( \begin{array}{c}
a_0 \\  \dots \\ (-1)^p a_p \end{array} 
 \right) 
= \left( \begin{array}{c}
-t_0 \\ \dots  \\  - \t_p \end{array} 
\right)
$$
This matrix is again a Vandermonde one, i.e., there exists always a unique 
solution $(a_0, \dots  a_p)$  (which
 are the coefficients of the quantum 
corrections to the Einstein equation) such that the  Newtonian limit of 
the corresponding gravitational field equation is a sum of Newtonian and 
Yukawa potential with prescribed lengths $l_i$. A more explicit form of the 
solution is given in the appendix.

\section{Discussion}

Let us give some special examples 
of the deduced formulas of the Newtonian 
limit of the theory described by the Lagrangian (1). 
If all the $a_i$ vanish we get of course the usual Newton theory
$$
\Phi = \frac{m}{r} \, , \quad 
    \Delta \Phi 
 = -4 \pi \d \, .
$$
$\Phi $   and $\Psi $ refer to
 the metric
 according to eq. (5). For
$ p = 0$ we 
get for $a_0 <0$
$$
\Phi =     \frac{m}{r} \left[
 1 + \frac{1}{3} \, e^{-r/\sqrt{-6 a_0}}
\right]
$$
(cf. STELLE 1978)  and  
$$
\Psi =     \frac{m}{r} \left[
 1 - \frac{1}{3} \, e^{-r/\sqrt{-6 a_0}}
\right]
$$
(cf. SCHMIDT 1986).
For $ a_0 > 0$  no Newtonian limit exists.

\bigskip

For $p = 1$, i.e., the theory following from sixth-order gravity
$$
{\cal L} = \left( R + a_0 R^2 +  a_1  R \Box R 
 \right) \sqrt{-g} \, , 
$$
we get
$$
\Phi =     \frac{m}{r} \left[
 1 + c_0  e^{-r/l_0} +  c_1  e^{-r/l_1}
\right]
$$
and
$$
\Psi =     \frac{m}{r} \left[
 1 - c_0 e^{-r/l_0} - c_1 e^{-r/l_1}
\right]
$$
where
$$
c_{0,1} = \frac{1}{6} \mp 
\frac{a_0}{2 \sqrt{9a_0^2 + 6 a_1}}
$$
and
$$
l_{0,1}= 
\sqrt{- 3 a_0 \pm
\sqrt{9 a_0^2 + 6a_1
}} \, .
$$
(This result is similar in structure but 
has different coefficients as 
in fourth-order gravity with 
included square of the Weyl tensor in the Lagrangian.)

\bigskip

The Newtonian limit for the degenerated 
case $l_0 = l_1$
 can be obtained by a limiting procedure as follows:
 As we already know $a_0 <0$,
 we can choose the length unit such that $a_0 = - 1/3$. 
The limiting case $9 a_0^2 + 6a_1 \to 0$ 
may be expressed by $a_1 = - 1/6 + \e^2$.  
After linearization in $\e$ we get:
$$
l_i = 1 \pm \sqrt{3/2} \, \e \, c_i   = 1/6 \pm 1/(6 
\sqrt 6 \e)
$$
and applying the limit $\e \to 0$
to the corresponding fields $\Phi $ and $\Psi$ we get 
\bea
\Phi  = m/r  \{1 + (1/3 + r/6) e^{-r} \}
\nonumber \\
 \Psi =    m/r  \{1 - (1/3 + r/6) e^{-r} \} \, . \nonumber 
\eea
For the general case $p > 1$, the potential is a complicated expression, 
but some properties are explicitly known, (these hold also for $p = 0, 1$).
 One gets
$$
\Phi  = m/r \, \left( 1 +  \sum_{i=0}^p \,   c_i  \exp (-r/l_i) \right)
$$
and
$$
\Psi  = m/r \, \left( 1 -  \sum_{i=0}^p \,   c_i  \exp (-r/l_i) \right)
$$
where $\sum c_i  = 1/3$; $\sum$ means 
$\sum_{i=0}^p$
 and $l_i$  and $c_i$
 are 
(up to permutation of indices) uniquely determined by the Lagrangian.

\bigskip

There exist some inequalities between the 
coefficients $a_i$, which must be fulfilled in 
order to get a physically acceptable
Newtonian limit. By this phrase we mean 
that besides the above conditions, additionally the fields $\Phi$  and
 $\Psi$  vanish for
$r \to \infty$
 and that the derivatives $d\Phi /dr$ and $d\Psi /dr$
 behave like $O(1/r^2)$. These inequalities 
express essentially the fact that
the $l_i$
 are real, positive, and different from each other. 
The last of these three conditions can be weakened by allowing the
 $c_i$  to be polynomials in $r$
 instead of being constants, cf. the example 
with $p = 1$ calculated above.

\bigskip

The equality $\sum c_i = 1/3$  means that the gravitational potential is 
unbounded and behaves (up to a factor 4/3) like the Newtonian 
potential for $r \approx  0$.
The equation $\Phi  + \Psi = 2m/r$
 enables us to rewrite the metric as
$$
ds^2 = (1 - 2 \theta) \left[
(1-    2m/r) dt^2
- (1  + 2m/r)    (dx^2 + dy^2 + dz^2) \right] \, ,
$$
which is the conformally transformed linearized
 Schwarzschild metric with the conformal factor
$1 - 2\theta$, where
$$
\theta  = \frac{m}{r} \sum c_i e^{-r/l_i}
$$
can be expressed as functional of 
the curvature scalar, this is the linearized
 version of 
the conformal transformation theorem, cf. SCHMIDT (1990).

For an arbitrary matter configuration the gravitational 
potential can be obtained by the usual integration procedure.

\section*{Appendix}
For general $p$ and characteristic lengths $l_i$ fulfilling 
$0 < l_0 < l_1 < \dots < l_p$  we write the Lagrangian as
\bea
{\cal L} =R - \frac{R}{6} \left[
(l_0^2 + \dots + l_p^2) R + (l_0^2 l_1^2 + l_0^2l_2^2
 \dots + l_{p-1}^2 l_p^2) \Box  R + \right.
\nonumber \\
\left. 
(l_0^2 l_1^2l_2^2 + \dots +  l_{p-2}^2 l_{p-1}^2
 l_p^2) \Box^2 R + \dots 
+ l_0^2 \cdot l_1^2 \cdot \dots \cdot l_p^2 \Box^p R
\right] \nonumber
\eea
the coefficients in front of $\Box^i R$ in this formula read
$$
\sum_{0 \le j_0 < j_1 < \dots < j_i \le p}
 \ \prod_{m=0}^i \, l^2_{j_m}
\, .
$$
Using this form of the Lagrangian, the gravitational potential of a 
point mass reads
\bea
\Phi =     \frac{m}{r} \left[
 1 + \frac{1}{3}
 \sum_{i=0}^p (-1)^{i+1}
 \prod_{j\ne i}
   \vert \frac{l_j^2}{l_i^2} -1 \vert ^{-1} \, 
  e^{-r/l_i}
\right] \, , 
\nonumber \\
\Psi =     \frac{m}{r} \left[
 1 - \frac{1}{3}
 \sum_{i=0}^p (-1)^{i+1}
 \prod_{j\ne i}
   \vert \frac{l_j^2}{l_i^2} -1 \vert ^{-1} \, 
  e^{-r/l_i}
\right] \, . 
\nonumber
\eea
For a homogeneous sphere of radius $r_0$  and mass $m$
 we get
$$
\Phi =     \frac{m}{r} \left[
 1 + \frac{1}{r_0^3}
 \sum_{i=0}^p
  e^{-r/l_i} \, l_i^2 \, \tilde c_i 
\left( 
r_0 \cosh (r_0/l_i) - l_i \sinh (r_0/l_i)
 \right) 
\right] \, , 
$$
where
$$
\tilde c_i = 
 (-1)^{i+1}
 \prod_{j\ne i}
   \vert \frac{l_j^2}{l_i^2} -1 \vert ^{-1}
\, .
$$

\section*{References}

\noindent 
BUCHDAHL, H.: 1951, Acta Math. {\bf 85}, 63.

\noindent 
DAUTCOURT, G.: 1964, Acta Phys. Polon. {\bf  25}, 637.

\noindent 
GERBAL, D., SIROUSSE-ZIA, H.: 1989,
 C. R. Acad. Sci. Paris, {\bf  309}, Serie 2, 353.

\noindent 
GOTTL\"OBER, S., SCHMIDT, H.-J., 
STAROBINSKY, A. A.: 1990, Class. Quantum Grav. {\bf  7},
893.

\noindent 
SANDERS, R. H.: 1984, Astron. Astrophys. {\bf  136}, L21.

\noindent 
SANDERS, R. H.: 1986, Astrophys. J. {\bf  154}, 135.

\noindent 
SCHMIDT, H.-J.: 1986, Astron. Nachr. {\bf  307}, 339.

\noindent 
SCHMIDT, H.-J.: 1990, Class. Quantum Grav. {\bf  7}, 1023.

\noindent 
STACEY, F. D. et a1.: 1981, Phys. Rev. D {\bf  23}, 2683.

\noindent 
STELLE, K.: 1978, Gen. Rel. Grav. {\bf  9}, 353.

\noindent 
STEPHANI, H.: Allgemeine Relativit\"atstheorie, Berlin 1977.

\noindent 
TEYSSANDIER, P.:   1989, Class. Quant. Grav. {\bf  6}, 219.

\noindent 
YUKAWA, H.: 1935, Proc. Math. Phys. Soc. Japan {\bf 17}, 48.

\bigskip

Remark added in proof: The cosmological 
consequences of these sixth-order field equations 
are recently discussed in A. Berkin 
and K. Maeda (Phys. Lett. B {\bf  245}, 348; 1990) and S. Gottl\"ober, 
V. M\"uller and H.-J. Schmidt (1990; preprint PRE-ZIAP 90-16).

\bigskip

{\it Received 1990 August 10}

\bigskip

\noindent 
{\small {  In this reprint done with the kind permission of the 
copyright owner 
we removed only obvious misprints of the original, which
was published in Astronomi\-sche Nachrichten:   
Astron. Nachr. {\bf 312} (1991) Nr. 2, pages 97 - 102.

\bigskip

\noindent 
  Authors's address  that time: 

\medskip

\noindent
I. Quandt, Humboldt-Universit\"at zu Berlin, 
Unter den Linden 6, O-1086    Berlin

\medskip

\noindent
H.-J. Schmidt,  
Zentralinstitut f\"ur  Astrophysik, 
O-1591 Potsdam--Babelsberg,  R.-Luxemburg-Str. 17a
}}
\end{document}